# Effect of Impurities Scattering Potential on NMR relaxation rate in impure d-wave superconductors

**P.Udomsamuthirun[1,2] and K.Meemon[1]**

(1) Department of Physics, Faculty of Science, Srinakharinwirot University, Bangkok 10110,Thailand. E-mail: udomsamut55@yahoo.com
(2) Thailand Center of Excellence in Physics, Huay Kaew Road, Chiang Mai, Thailand, 50200

**Abstract**

The purpose of our research is to study the nuclear spin lattice relaxation rate of impure d-wave superconductors. We use the Green's function method to derive the approximation equation of density of states including the impurity scattering potential. We can get the analytic equation of the nuclear spin lattice relaxation rate that contained the impurity scattering potential in case of weak scattering potential and strong scattering potential in the simple form as the power series of $\Delta(T)$ and $T$. The numerical calculations show that there is coherence peak in the weak impurity scattering potential but there is no peak in the strong impurity scattering potential.

**Keywords:** Impure d-wave Superconductors, Density of States, The nuclear spin lattice relaxation rate

**PACS :** 74.20.Fg, 76.60.-k, 74.50

## 1. Introduction

The study of unconventional superconductors has attracted the most interest of scientist in field of condensed matter physics. NMR experiment is one of the probe to clarify that is the conventional or unconventional superconductors. It is well known that the nuclear spin lattice relaxation rate $T_1^{-1}$ in s-wave superconductor has a Hebel-Slichter[1] peak while unconventional superconductors do not exhibit this feature. Leadon and Suhl [2] proposed $T_1^{-1}$ equation that proportional to the probability of flipping of the nuclear spin writing in term of normal and anomalous temperature Green's function. Ishida at al. [3] found that $T_1^{-1}$ is proportional to $T$ at low temperature that affect by the impurities in the superconductors. Hasehawa [4] show that the density of state with intersecting line nodes in the gap function can show the power law in relaxation rate $T_1^{-1}$. Choi[5] studied the thermodynamic properties of impure superconductors within the phonon-mediated Eliashberg formulism. The $\frac{1}{TT_1}$ is also calculated. Band et al.[6] considered a general d-wave superconducting gap function in the presence of coexisting antiferromagnetic long rang order. They calculated numerically the $T_1^{-1}$ of gap in the presence of impurities. However, the study of influence of impurities on the electronic states in the conventional and unconventional superconductors is reviewed by Balatsky, Vekhter, and Jian-Xin Zhu[7]. Matsumoto[8] studied the impurity nuclear spin relaxation rate in unconventional superconductor. Parker and Hass[9] demonstrate that within the BCS theory the nuclear spin lattice relaxation rate $T_1^{-1}$ should show a small peak below $T_c$ for unconventional superconductor.

In this research, we calculate to find the simple analytic nuclear spin lattice relaxation rate $T_1^{-1}$ equation for impure d-wave superconductors. The Green's function is used to find the density of state and we substitute it into $T_1^{-1}$ equation [2,8]. And the numerical calculations are shown.

## 2. Model and Calculations

We begin with the BCS Hamiltonian of superconductor,

$$H = \sum_{k\sigma}\varepsilon_k a_{k\sigma}^+ a_{k\sigma} + \sum_{kk'} V_{kk'} a_{k\uparrow}^+ a_{-k\downarrow}^+ a_{-k'\downarrow} a_{k'\uparrow} \qquad (1)$$

Here the operator $a_{k\sigma}^+$ ($a_{k\sigma}$) creates(annihilates) an electron with the wave vector $k$ and the spin projection on z-axis σ = ↑ or ↓, $\varepsilon_k = \xi_k - \mu$ is the (spin independent) quasi-particle energy measured from the chemical potential. $V_{kk'}$ is the BCS pair potential.

For the formal procedure of calculation, we can get matrix of non-perturbed Green function in the Nambu representation as

$$G_0(i\omega_m, r, r') = -\frac{1}{\Omega}\sum_k e^{ik\cdot(r-r')} \frac{i\omega_m + \varepsilon_k \tau_3 + \Delta_k \tau_1}{\omega_m^2 + \varepsilon_k^2 + \Delta_k^2} \qquad (2)$$

where $\omega_m$ is the Matsubara frequency and $\Delta_k$ is the order parameter. $\tau_i$ (i=1,2,3) is the Pauli matrix. $\Omega$ is the volume of the system.

As usual the elastic scattering problem, The Matsubara Green function is

given by

$$G(i\omega_m,r,r') = G_0(i\omega_m,r,r') + G_0(i\omega_m,r,0)U_0\tau_3 \frac{1}{1-G_0(i\omega_m,0,0)U_0\tau_3} G_0(i\omega_m,0,r')$$

(3)

Where a single impurity is located at the origin of the coordinate. $U_0$ represents the strength of the short-range impurity potential.

And the density of state is defined as

$$N_{imp}(E) = -\frac{1}{\pi N_0}\text{Im}[G_{11}(i\omega_m \to E+i\delta,0,0)] \quad (4)$$

where $N_0$ is the density of state at Fermi energy in normal state. $\delta$ is a small positive number.

Substitution Eq.(3) into Eq.(4), the density of state of impure superconductors is

$$N_{imp}(E) = \frac{N_d(E)}{1+u_0^2 N_d^2(E)} \quad (5)$$

Where $N_d(E) = \frac{E}{\sqrt{E^2-\Delta_k^2}}$ is the BCS density of state and $u_0 = \pi N_0 U_0$.

The $T_1^{-1}$ equation for superconductors defined by [2,8] is used to study the effect of impurity on the nuclear spin lattice relaxation rate. As $T_1^{-1}$ is proportional to the probability $W_{n\to m}$ of flipping the nuclear spin from the state $|n>$ to the $|m>$, it shows the relation below

$$T_1^{-1} = \frac{2}{\pi}(\frac{4\pi}{3})^2(\gamma_e\gamma_n)^2 W \quad (6)$$

Here $W$ is the nuclear spin flip transition probability,

$W = \int dE \frac{a_{11}(E)a_{22}(-E) - a_{12}(E)a_{21}(-E)}{1+\cosh(E/T)}$. $\gamma_e$ and $\gamma_n$ are the gyromagnetic ratios for the electron and nucleon. $a_{ij}$ represents the matrix element of the Green function. For d-wave pairing, we can get $N_{imp}(E) = \frac{a_{11}(E)}{N_0} = \frac{a_{22}(-E)}{N_0}$, that is the dimensionless density of states and $a_{12}$ and $a_{21}$ are zero, since pair electron cannot possess the same position. $W$ is then expressed by the following simple form below

$$W = N_0^2 \int_0^{\omega_D} \text{sec}h^2(\frac{E}{2T})N_{imp}^2(E)dE \quad (7)$$

In $d_{x^2-y^2}$–wave superconductors, the order parameter is $\Delta_k(T) = \Delta(T)\sin 2\theta$ or $\Delta_k(T) = \Delta(T)\cos 2\theta$ in 2-dimensional system. $\Delta_k(T)$ is temperature and angular dependence order parameters and $\Delta(T)$ is the temperature dependent order parameter. The angular dependent functions are $\cos 2\theta$ and $\sin 2\theta$ where $\theta$ is the angle of the Fermi wave vector measured from the $k_x$ axis. The density of state, Eq.(5) can be rewrite as

$$N_{imp}(E,\theta) = \frac{N_d(E,\theta)}{1+u_0^2 N_d^2(E,\theta)} \quad (8)$$

where $N_{imp}(E) = \int d\Omega N_{imp}(E,\theta)$.

Substitution Eq.(8) into Eqs.(6) and (7), we can get the effect of impurity on the nuclear spin lattice relaxation rate $T_1^{-1}$. However, this function can not be calculated analytically so we will consider two case, weak scattering potential and strong scattering potential.

In the weak scattering potential, the $u_0 N_d(E) << 1$ is assumed. Using approximation and average over angular Fermi surface. We can get [10]

$$N_{imp}(E) \approx \frac{2}{\pi} K(\frac{\Delta}{E}) - \frac{2}{\pi} u_0^2 (1+\frac{\Delta^2}{E^2}) E(\frac{\Delta}{E}) \quad \text{for } E > \Delta$$

$$\approx \frac{2}{\pi} \frac{E}{\Delta} K(\frac{E}{\Delta}) \quad \text{for } E < \Delta \quad (9)$$

Here $K(E)$ and $E(E)$ are the complete elliptic integral of first and second kind.

Substituting Eq.(9) into Eq.(7), and taking $\sec h^2 x = 1 - \tanh^2 x$ and using the approximation as $\tanh x \approx x$, for $x < 1$ and $\tanh x \approx 1$, for $x > 1$. Eq.(7) must be separated into two case; $\Delta < 2T$ and $\Delta > 2T$.

The nuclear spin flip transition probability $W$ for the $\Delta < 2T$ case is

$$\frac{W}{T} = 2N_0^2 \left\{ 0.6680 + d_1 + 0.3504 \frac{\Delta}{2T} - 0.4534(\frac{\Delta}{2T})^3 + (-1.3400 - d_2 + 0.3060\frac{\Delta}{2T})u_0^2 + (0.6720 + d_3 - 0.00961\frac{\Delta}{2T})u_0^4 \right\}$$

(10)

and for $\Delta > 2T$ case is

$$\frac{W}{T} = 2N_0^2 \left\{ 0.1333(\frac{2T}{\Delta})^2 + 0.0286(\frac{2T}{\Delta})^4 + e_1(\frac{2T}{\Delta})^{\beta_1} + f_1 + (e_2(\frac{2T}{\Delta})^{\beta_2} + f_2)u_0^2 + (e_3(\frac{2T}{\Delta})^{\beta_3} + f_3)u_0^4 \right\}$$

(11)

Here $x = \frac{E}{\Delta}$ and the higher order terms are defined as $d$ and $e(\frac{2T}{\Delta})^{\beta} + f$ where $d, e, f, \beta$ are constant.

There are 3 conditions for Eq.(10) and Eq.(11) ; 1) At $T=T_c, \frac{W_n}{W_s}\big|_{T=T_c} = 1$. 2) At $T \to 0, \frac{W_n}{W_s}\big|_{T=T_c} = 0$. And 3) the functions and their first derivative must be continuous at $\Delta = 2T$. After taking 3 conditions, we can get the nuclear spin lattice relaxation rate for $\Delta(T) < 2T$ as

$$\frac{1}{T_1} = \frac{2}{\pi}(\frac{4\pi}{3})^2 (\gamma_e \gamma_n)^2 N_0^2 T \left\{ 1 + 0.3504\frac{\Delta}{2T} - 0.4534(\frac{\Delta}{2T})^3 + (-2+0.360\frac{\Delta}{2T})u_0^2 + (3 - 0.00961\frac{\Delta}{2T})u_0^4 \right\}$$

(12)

, and for $\Delta(T) > 2T$ as

$$\frac{1}{T_1} = \frac{2}{\pi}(\frac{4\pi}{3})^2 (\gamma_e \gamma_n)^2 N_0^2 T \left\{ 0.1333(\frac{2T}{\Delta})^2 + 0.0286(\frac{2T}{\Delta})^4 + 0.7351(\frac{2T}{\Delta})^{0.86} - 1.694(\frac{2T}{\Delta})^{0.18} u_0^2 + 2.994(\frac{2T}{\Delta})^{0.0032} u_0^4 \right\}$$

(13)

In the strong scattering potential, the $u_0 N_d(E) >> 1$ is assumed. The density of state is

$$N_{imp}(E) \approx \frac{2}{\pi u_0^2} E(\frac{\Delta}{E}) - \frac{2}{3\pi u_0^4} \left\{ (\frac{4E^2 - 2\Delta^2}{E^2}) E(\frac{\Delta}{E}) - (\frac{E^2 - \Delta^2}{E^2}) K(\frac{\Delta}{E}) \right\} \quad \text{for } E > \Delta$$

$$\approx \frac{2}{\pi u_0^2} \frac{\Delta}{E} \left\{ E(\frac{E}{\Delta}) + (\frac{E^2}{\Delta^2} - 1)K(\frac{E}{\Delta}) \right\}$$

$$- \frac{2}{3\pi u_0^4} \left\{ (\frac{4E^2\Delta - 2\Delta^3}{E^3}) \left[ E(\frac{E}{\Delta}) + (\frac{E^2 - \Delta^2}{\Delta^2})K(\frac{E}{\Delta}) \right] - (\frac{E^2 - \Delta^2}{E\Delta})K(\frac{E}{\Delta}) \right\} \quad \text{for } E < \Delta \tag{14}$$

We can get the nuclear spin lattice relaxation rate for $\Delta(T) < 2T$ as

$$\frac{1}{T_1} = \frac{2}{\pi} (\frac{4\pi}{3})^2 (\gamma_e \gamma_n)^2 N_0^2 T \left\{ \frac{1}{u_0^4} \left[ 1 - 1.2759(\frac{\Delta}{2T}) - 0.0657(\frac{\Delta}{2T})^3 + 0.3333(\frac{\Delta}{2T})^5 \right] \right.$$

$$+ \frac{1}{u_0^6} \left[ -2 + 3.171\frac{\Delta}{2T} + 0.0930(\frac{\Delta}{2T})^3 - 1.0219(\frac{\Delta}{2T})^5 \right]$$

$$\left. + \frac{1}{u_0^8} \left[ 3 - 1.8466\frac{\Delta}{2T} - 0.0329(\frac{\Delta}{2T})^3 - 0.4272(\frac{\Delta}{2T})^5 \right] \right\} \tag{15}$$

,and for $\Delta(T) > 2T$ as

$$\frac{1}{T_1} = \frac{2}{\pi} (\frac{4\pi}{3})^2 (\gamma_e \gamma_n)^2 N_0^2 T \left\{ \frac{1}{u_0^4} \left[ 0.0333(\frac{2T}{\Delta})^2 + 0.0036(\frac{2T}{\Delta})^4 - 0.0451(\frac{2T}{\Delta})^6 \right] \right.$$

$$+ \frac{1}{u_0^6} \left[ -0.0500(\frac{2T}{\Delta})^2 - 0.0045(\frac{2T}{\Delta})^4 + 0.2965(\frac{2T}{\Delta})^6 \right]$$

$$\left. + \frac{1}{u_0^8} \left[ 0.0188(\frac{2T}{\Delta})^2 + 0.0013(\frac{2T}{\Delta})^4 + 0.6732(\frac{2T}{\Delta})^6 \right] \right\} \tag{16}$$

Here the higher order terms are defined as $h(\frac{\Delta}{2T})^5 + g$ and $k(\frac{2T}{\Delta})^6 + m$ where $g, h, m, k, h$ are constant that can find by taking the boundary conditions.

For pure d-wave superconductors, $u_0 N_d(E) = 0$. Our calculations agree with Reference [11].

The parameter we need for $T_1^{-1}$ calculation is the temperature dependent order parameter, $\Delta(T)$. For more accuracy, the same process as Ref.[12-13] is used to find $\Delta(T)$ within the determined parameters $\Delta(0)$, $T_c$ and $\omega_D$. The temperature dependence energy gap can get from the equation

$$\int_{-\omega_D}^{\omega_D} d\varepsilon \frac{\tanh(\varepsilon/2T)}{\varepsilon} = \frac{1}{2\pi} \int_0^{2\pi} d\theta \int_{-\omega_D}^{\omega_D} d\varepsilon \frac{N_{imp}(\varepsilon, \Delta(T))\tanh(\sqrt{\varepsilon^2 + \Delta_k^2(T)}/2T)}{\sqrt{\varepsilon^2 + \Delta_k^2(T)}} \tag{17}$$

with zero temperature gap as

$$\int_{-\omega_D}^{\omega_D} d\varepsilon \frac{\tanh(\varepsilon/2T)}{\varepsilon} = \frac{1}{2\pi} \int_0^{2\pi} d\theta \int_{-\omega_D}^{\omega_D} d\varepsilon \frac{N_{imp}(\varepsilon, \Delta(0))}{\sqrt{\varepsilon^2 + \Delta_k^2(0)}} \tag{18}$$

### 3. Results and Discussions

In our calculation, we can get the nuclear spin lattice relaxation rate $T_1^{-1}$ that contained the impurity scattering potential in case of weak scattering potential and strong scattering potential in the simple form as the power series of $\Delta(T)$ and T. The effects of weak scattering potential are calculated by using Eqs.(12-13) and Eqs.(17-18) as shown in Figure.1.1 and Figure 1.2.

Because most of the experimental data of $T_1^{-1}$ are shown in log-log plot so we show our results in both linear-linear plot and log-log plot. In linear-linear plot, we can see the coherence peak clearly. However, the $T_1^{-1} \alpha T^3$ in low temperature range can see easily in the log-log plot.

We find that the curve of $T_1^{-1}$ and $(TT_1)^{-1}$ are also shown the coherence peak below $T_c$. Within our numerical accuracy, the parameters we used have no effect on $(TT_1)^{-1}$ so we get the same result. And Eqs.(15-16) and Eqs.(17-18) are used for numerical calculations of $T_1^{-1}$ and $(TT_1)^{-1}$ in strong scattering potential. The effect of strong scattering potentials are shown in Figure.2.1 and Figure 2.2. We find that the curve of $T_1^{-1}$ and $(TT_1)^{-1}$ do not show the coherence peak below $T_c$. Within our numerical accuracy, the parameters we used are not effect to $(TT_1)^{-1}$ so we get the same result.

There are many experiments that study on the temperature dependence of $T_1^{-1}$ of unconventional superconductors. The $T_1^{-1}$ of unconventional superconductors Pr(Os$_{1-x}$Ru$_x$)$_4$Sb$_{12}$ were studied by Kotegawa et al.[14] for x=0, Nishiyama et al.[15] for x=0.1,0.2 and Yogi et al.[16] for x=1.They found coherence peak at x=0,0.1,0.2 and no peak at x=1. Nishiyama, Inada and Zheng[17] found coherence peak in unconventional superconductors in Li$_2$Pd$_3$B and the coherence peak in YBCO materials was found by Ishida et al. [3].

However, the other kinds of unconventional superconductors as heavy-fermion superconductors that can not find the coherence peak. In heavy-fermion superconductors, such as $CeCu_2Si_2$, $CeRhIn_5$, $CeIn_3$ [18], $CeCoIn_5$ [19] and else, there are the magnetic order in below $T_c$ region. The random magnetic orders can be considered as the impurities so they can decrease the critical temperature of superconductors[20]. However, to find the $T_1^{-1}$ equation contained the non-random magnetic orders, the Hamiltonian of this system must be changed. So our model can not cover these materials. In our view, every kind of superconductors can apply to use our model if we know the Green function of those systems. .

For our results, we can conclude that the coherence peak of the nuclear spin lattice relaxation rate $T_1^{-1}$ can be found in pure d-wave superconductors that agree with some experiments. There is the other factor that effect on the measurement of coherence peak such as magnetic orders[6] that we do not include in our calculation.

**4.Conclusions**

In this research, we show the analytical nuclear spin lattice relaxation rate $T_1^{-1}$ equation for impure d-wave superconductor in the simple form contained the impurity scattering potential in case of weak scattering potential and strong scattering potential. The numerical calculation of $T_1^{-1}$ and $(TT_1)^{-1}$ are shown. We can conclude that the nuclear spin lattice relaxation rate $T_1^{-1}$ of d-wave superconductors can or can not show a small peak below $T_c$ depending on strength of scattering potential.

**Acknowledgement**
The author would like to thank Professor Dr.Suthat Yoksan for the useful discussions and also thank Srinakharinwirot University and ThEP Center for the financial support.

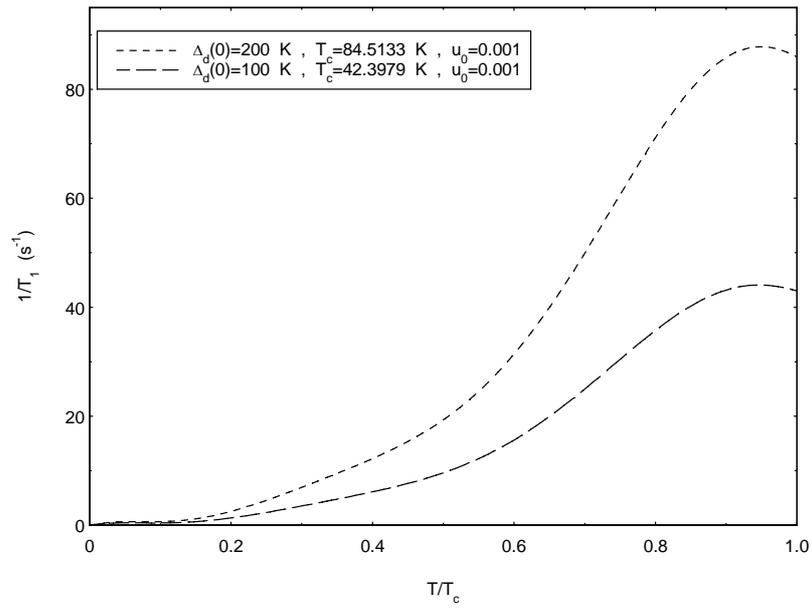

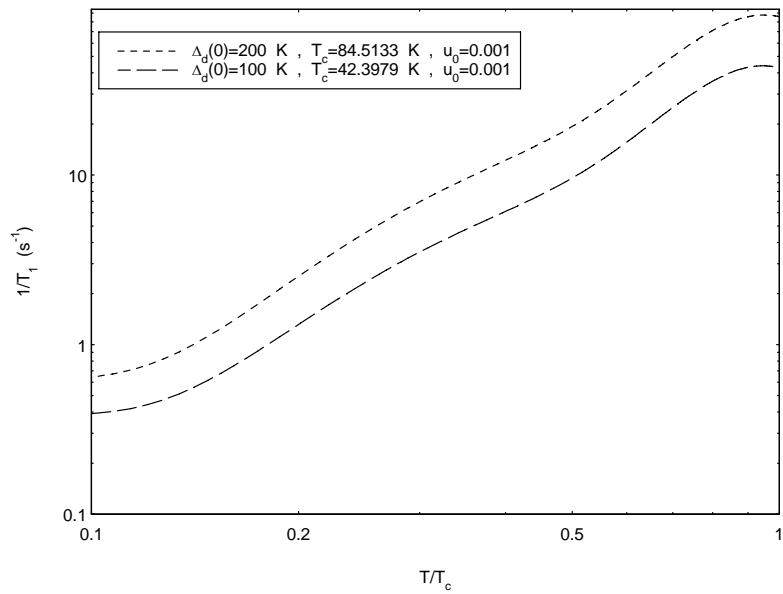

Figure 1.1 The effect of weak scattering potential on $T_1^{-1}$ are shown in linear-linear plot (a) and log-log plot (b) .

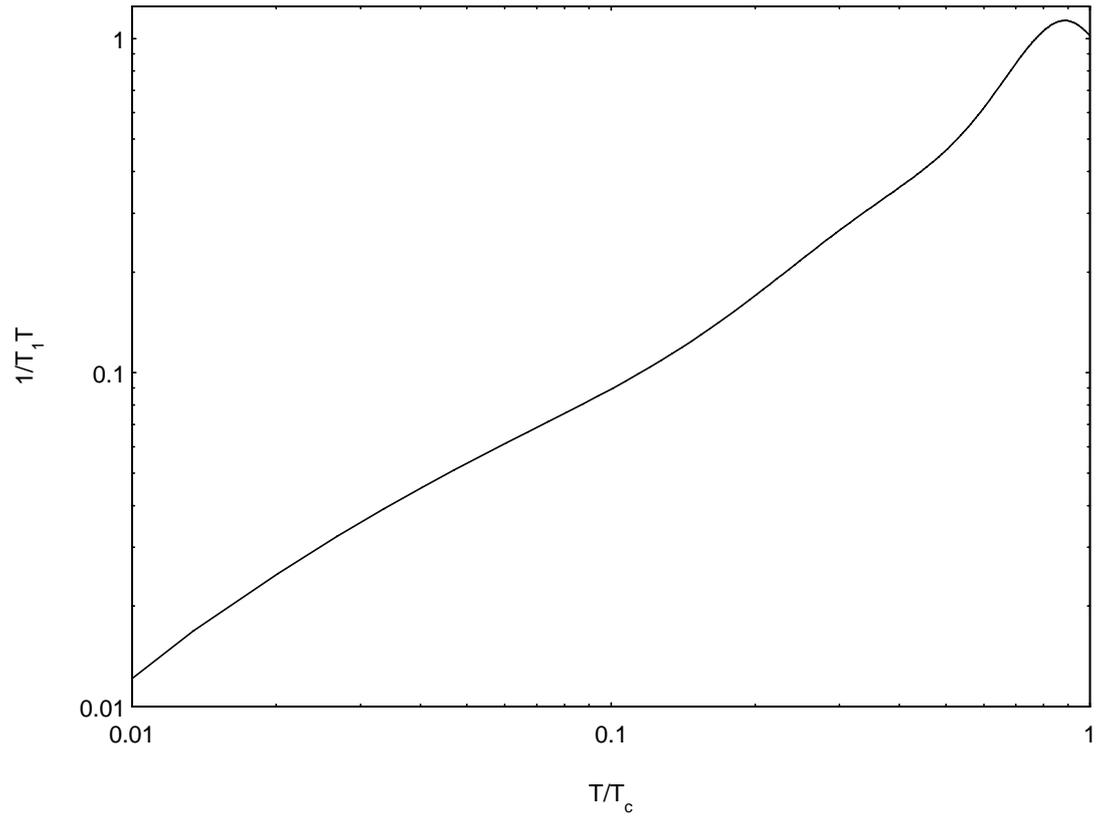

Figure 1.2 Within the results in Figure 1.1 ,the effect of weak scattering potential on $(TT_1)^{-1}$ are shown in log-log- plot that coincided within numerical accuracy.

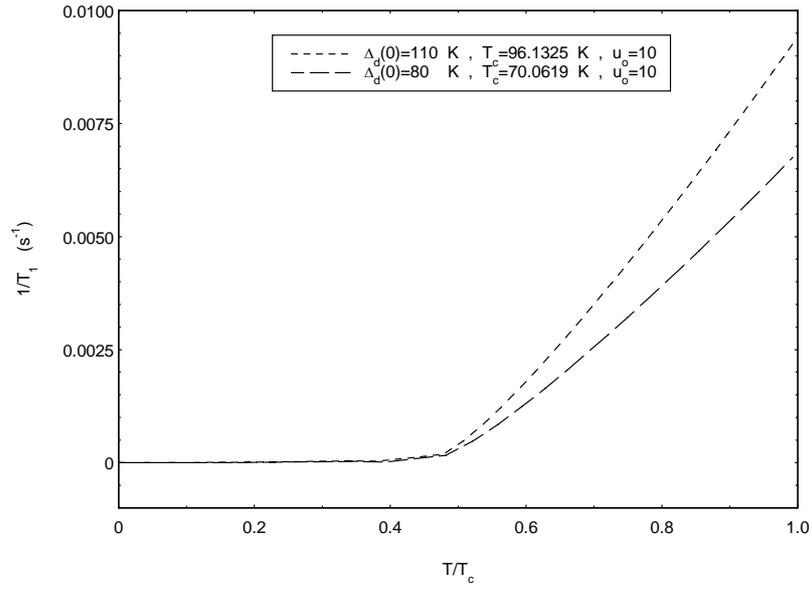

(a)

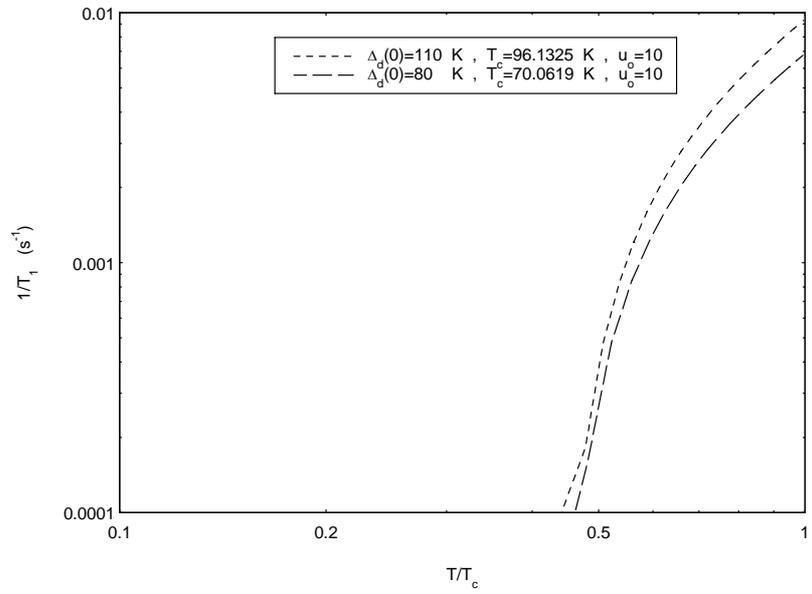

(b)

Figure 2.1 The effect of strong scattering potential on $T_1^{-1}$ are shown in linear-linear plot (a) and log-log plot (b) .

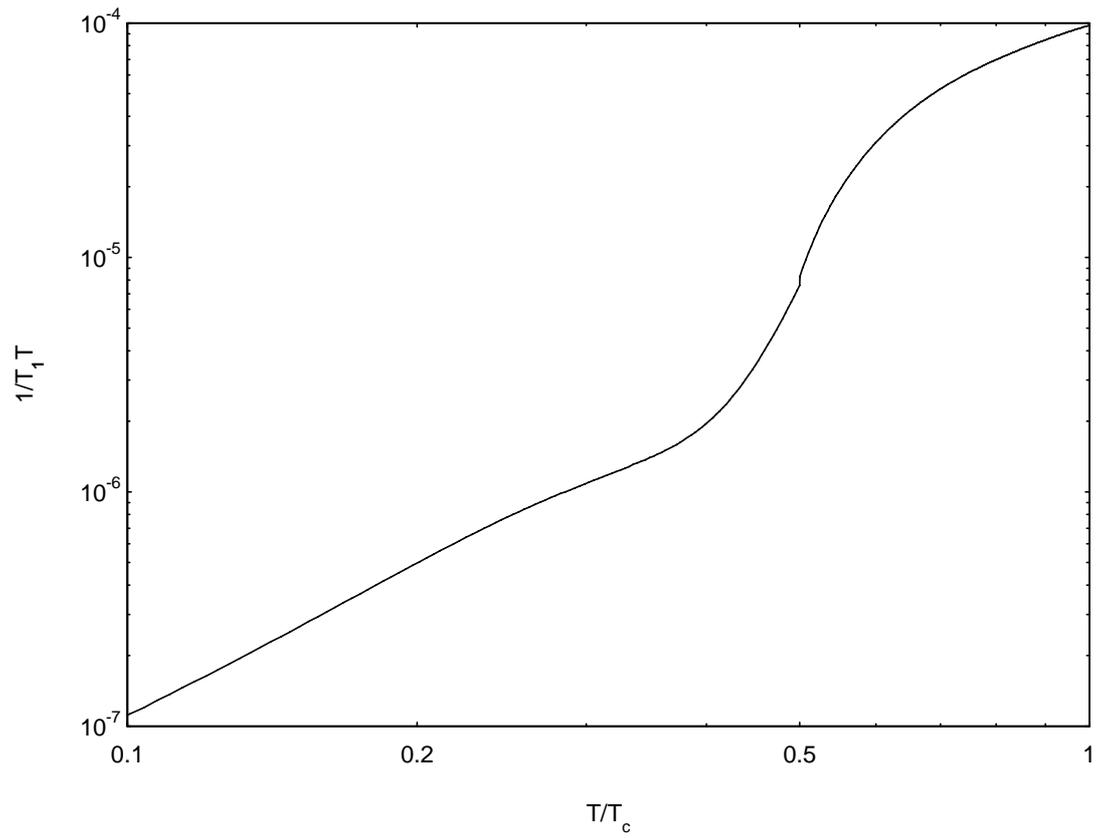

Figure 2.2 Within the results in Figure 2.1,the effect of weak scattering potential on $(TT_1)^{-1}$ are shown in log-log- plot that coincide within numerical accuracy.